\documentclass[aps,prl,reprint,superscriptaddress]{revtex4-2}
\usepackage{amsfonts,amssymb,amsmath}
\usepackage{enumerate}
\usepackage[sort&compress]{natbib}
\usepackage{hyperref}
\usepackage{graphicx}

\begin{document}

%\preprint{}

\title{Equilibrium and nonequilibrium description of negative temperature states in a one dimensional lattice using a wave kinetic approach}


\author{M. Onorato}
%\thanks{}
\affiliation{Dipartimento di Fisica, Universit\`a degli Studi di Torino, 10125 Torino, Italy}
\affiliation{Istituto Nazionale di Fisica Nucleare, INFN, Sezione di Torino, 10125 Torino, Italy}
\author{G. Dematteis}
%\thanks{}
\affiliation{Department of Mathematical Sciences, Rensselaer Polytechnic Institute, Troy, NY 12180, US}
\author{D. Proment}
%\thanks{}
\affiliation{School of Mathematics, University of East Anglia, Norwich Research Park, NR47TJ Norwich, UK}
\author{A. Pezzi}
%\thanks{}
\affiliation{Dipartimento di Fisica, Universit\`a degli Studi di Torino, 10125 Torino, Italy}
\author{M. Ballarin}
%\thanks{}
\affiliation{Dipartimento di Fisica, Universit\`a degli Studi di Torino, 10125 Torino, Italy}
\author{L. Rondoni}
%\thanks{}
\affiliation{Dipartimento di Scienze Matematiche, Politecnico di Torino, I-10129 Torino, Italy}
\affiliation{Istituto Nazionale di Fisica Nucleare, INFN, Sezione di Torino, 10125 Torino, Italy}


\date{\today}

\begin{abstract}
We predict  negative temperature states in the Discrete Nonlinear Sch\"odinger (DNLS) equation as exact solutions of the associated Wave Kinetic equation. Within the wave kinetic approach, we define an entropy that results monotonic in time and reaches a stationary state, that is consistent with classical  equilibrium statistical mechanics.
We also perform a detailed analysis of the fluctuations of the actions at fixed wave numbers around their mean values. We give evidence that such fluctuations relax to their equilibrium behaviour on a shorter time scale than the one needed for the spectrum to reach the equilibrium state. Numerical simulations of the DNLS equation are shown to be in agreement with our theoretical results.  The key ingredient for observing negative temperatures in lattices characterized by two invariants is the boundedness of  the dispersion relation.
\end{abstract}

%\keywords{}

\maketitle


\section{Introduction}
Negative temperatures have been the subject of intensive studies since they were conceived in the study of  point vortices \cite{onsager1949statistical}  and the subsequent experimental work in  \cite{purcell1951nuclear}, where negative temperatures in nuclear spin systems were observed. More recently, negative temperatures were also observed in ultra-cold quantum systems \cite{braun2013negative}; moreover, the predictions of Onsager on point vortices have been verified experimentally \cite{johnstone2019evolution,gauthier2019giant}. In thermodynamics the requirement for observing negative temperatures is that the entropy $S$ does not increase monotonically  with energy. Indeed, if the entropy  is a continuous function of the energy, $E$, reaching a maximum away from its domain boundary, negative temperatures are expected. This  descends from the thermodynamic definition of temperature, $T=(\partial S/\partial E)^{-1}$.  As expressed in \cite{ramsey1956thermodynamics}, ``the assumption of monotonic increase of the entropy with the energy is not essential to the development of thermodynamics...''. Despite controversies and criticisms to the existence of negative temperatures related to the definition of the entropy \cite{dunkel2014consistent,calabrese2019origin}, negative temperatures are now well accepted by a vast community and the observations in experiments appear to be reliable \cite{frenkel2015gibbs,buonsante2016dispute,puglisi2017temperature,cerino2015consistent,abraham2017physics,baldovin2021statistical}. 

Here, we present an approach to the theory of negative temperatures in weakly  anharmonic lattices based on the so called Wave Kinetic (WK) equation \cite{nazarenko2011wave,zakharov2012kolmogorov}, i.e. an equation that, in analogy with the Boltzmann equation for particles, describes the mesoscale dynamics of a system of interacting waves. The WK equation can be derived in a systematic way from deterministic (microscopic) weakly nonlinear and dispersive wave systems \cite{nazarenko2011wave,zakharov2012kolmogorov,onorato2020astraighforward}. It has been applied to a variety of fields such as nonlinear optics \cite{picozzi2014optical}, surface gravity waves \cite{zakharov1966energy,hasselmann1962non}, Bose-Einstein condensation \cite{nazarenko2006wave,proment2009quantum}, gravitational waves \cite{galtier2017turbulence}, vibrations in anharmonic lattices  \cite{lukkarinen2008anomalous,onorato2015route}. In this Report, we will consider the Discrete Nonlinear Schr\"odinger (DNLS) equation as the starting point, and build its thermodynamic properties in the limit of small nonlinearity, passing through a mesoscopic description via the WK equation.

The DNLS equation, \cite{kevrekidis2009discrete}, as its continuous version, is a universal model; it describes  the propagation of optical waves in a waveguide array or a Bose-Einstein condensate in a periodic optical lattice. Differently from other discrete systems, like the 
Fermi-Pasta-Ulam-Tsingou (FPUT) lattice \cite{fermi1955alamos}, the DNLS equation has two conserved quantities, i.e the Hamiltonian and the total number of particles (two conservation laws are a fundamental ingredient for observing negative temperatures). A number of previous studies \cite{rasmussen2000statistical,rumpf2008transition,rumpf2009stable,iubini2012nonequilibrium,iubini2013discrete,rumpf2007growth,rumpf2004simple}
have discussed  the statistical mechanics of the DNLS equation. The main idea   in \cite{rasmussen2000statistical} is that the negative temperatures in DNLS are associated with the emergence of high amplitude localized structures or discrete breathers  \cite{flach1998discrete} in the strongly nonlinear regime. Developments using the microcanonical ensemble can be found in \cite{gradenigo2019localization}. In the field of nonlinear optics some interesting work has been done at equilibrium for a finite number of modes, 
see \cite{parto2019thermodynamic,makris2020statistical,wu2019thermodynamic,wu2020entropic}.  Our approach, being based on a theory that makes use of the random phase approximation both for positive and negative temperatures, cannot be applied in the presence of coherent structures such as solitons or breathers. Negative temperatures will reveal themselves as  localized Fourier energy spectrum in the high wave number region.

\section{The Wave Kinetic theory for the DNLS equation}
The DNLS equation reads
\begin{equation}
i \dot \psi_m+(\psi_{m+1}+\psi_{m-1}-2\psi_{m})+\nu |\psi_m|^2\psi_m=0,
\label{eq:dnls}
\end{equation}
where $\psi_n$ is the complex amplitude of the oscillator at site $m$, with $m=1,2,...,M$ and 
$\nu$ is an anharmonic parameter that weighs the nonlinearity of the system. The DNLS equation has two conserved quantities:
\begin{equation}
\begin{split}
&\mathrm{H}=\sum_{m=1}^{M} \left(|\psi_{m+1}-\psi_{m}|^2-\frac{1}{2}\nu |\psi_{m}|^4\right),\\
&\mathrm{N}=\sum_{m=1}^{M} |\psi_{m}|^2\,.
\end{split}
\end{equation}
which are the Hamiltonian and the total conserved norm of the DNLS equation, respectively. 

In our work we will use periodic boundary conditions and, using the following convention for the Discrete Fourier Transforms,
\begin{equation}
\psi_m=\sum_{k=1}^{M}\hat \psi_k e^{i2\pi km/M },\;\;\; \hat\psi_k=\frac{1}{M}\sum_{n=1}^{M}\psi_m e^{-i2\pi km/M },
\end{equation}
we write the equation in Fourier space as
\begin{equation}
i \dot {\hat \psi}_{k_1}=\omega_{k_1} \hat\psi_{k_1}-\nu\sum_{k_2,k_3,k_4} \hat \psi_{k_2}^*\hat \psi_{k_3}\hat \psi_{k_4}
\delta_{12}^{34},
\label{eq:dnlsk}
\end{equation}
where   $\omega_{k}=4 \sin^2(\pi k/M)$  and $\delta_{12}^{34}=\delta_{k_1+k_2,k_3+k_4}$ is the Kronecker $\delta$ that accounts for Umklap processes, i.e. $k_1+k_2=k_3+k_4, \mod M$. 
The Hamiltonian in Fourier space takes the following form:
\begin{equation}
\frac{\mathrm{H}}{M}=\sum_{k_1}\omega_{k_1} |\hat \psi_{k_1}|^2-\frac{1}{2}\nu\sum_{k_1k_2,k_3,k_4} \hat \psi_{k_1}^*\hat \psi_{k_2}^*\hat \psi_{k_3}\hat \psi_{k_4}
\delta_{12}^{34},
\label{eq:ham_fourier}
\end{equation}

By using the following transformation
$\hat \psi_k=\sqrt{I_k} \exp(-i \theta_k)$, the equation can be written in angle-action variables:

\begin{equation}\label{eq:40}
\begin{split}
&\frac{d I_{k_1}}{d t} =-2 \nu\sum  \sqrt{I_{k_1}I_{k_2}I_{k_3}I_{k_4}} \sin(\Delta \theta_{12}^{34})\delta_{12}^{34}\,, \\
&\frac{d \theta_{k_1}}{d t} = \omega_{k_1} - \nu\sum \sqrt{\frac{I_{k_2}I_{k_3}I_{k_4}}{I_{k_1}}} \cos(\Delta \theta_{12}^{34})\delta_{12}^{34}\,,
\end{split}
\end{equation}
with $ \Delta \theta_{12}^{34}:=\theta_{k_1}+\theta_{k_2}-\theta_{k_3}-\theta_{k_4}$.
Assuming that $\nu\ll1$, we expand the action-angle variables in powers of $\nu$; we then assume that the initial angles (or phases) are independent random variables uniformly distributed in the $[0,2\pi)$ interval. A key step consists in taking the large box limit, which implies taking $M\rightarrow\infty$, thus making the Fourier modes dense in the interval $[0,2\pi)$. The  Wave Kinetic equation (Boltzmann equation for phonons)  can then be obtained (see \cite{onorato2020astraighforward} for details on the derivation):
\begin{equation}
\begin{split}
  \frac{dn_{k_1}}{d \tau} =\xi_{k_1}-\gamma_{k_1} n_{k_1}
 \label{eq:discretekinetic_avI1}
 \end{split}
 \end{equation}
with
\begin{equation}
\begin{split}
&\xi_{k_1} =4 \pi \nu^2\int_0^{2\pi} n_{k_2}n_{k_3}n_{k_4}
\delta({\Delta \omega_{12}^{34}} )\delta_{12}^{34} dk_{234} \\
& \gamma_{k_1}=- 4 \pi \nu^2\int_0^{2\pi} \left(n_{k_3}n_{k_4}  -n_{k_2}n_{k_3} - n_{k_2}n_{k_4}   \right)\times\\
&\delta({\Delta \omega_{12}^{34}} )\delta_{12}^{34}dk_{234},
\end{split}
\end{equation}
where $\Delta \omega_{12}^{34}=\omega_{k_1}+\omega_{k_2}-\omega_{k_3}-\omega_{k_4}$, 
$k$ is a continuous variable in the $[0,2\pi]$ interval, $dk_{234}=dk_2 dk_3 dk_4$,  $\omega_k=4 \sin(k/2)^2$, $n_{k}=n(k,t)=\langle {I_k} \rangle M/2\pi$
is the wave action spectral density, and  $\langle \, \cdot \, \rangle$ is performed over the initial random phases and independent actions. Strictly speaking, the WK equation is valid under the assumption that random phases and amplitudes persist over time \cite{chibbaro20184}.
 Besides the evolution equation for the spectral density function, using the same approximations and tools, it is also possible to derive an evolution equation for the second moment, $\Lambda_k=\langle I_k^2 \rangle  (M/2\pi)^2$, which reads:
\begin{equation}
\begin{split}
& \frac{d\Lambda_{k_1}}{d \tau} =4 n_{k_1}\xi_{k_1} -2 \gamma_{k_1}  \Lambda_{k_1} .
\label{eq:variance}
\end{split}
\end{equation}
Such an equation, see \cite{nazarenko2011wave,tanaka2013numerical}, describes the fluctuations of the wave action density at fixed wave number.
The solution of the Cauchy problem for  the coupled system (\ref{eq:discretekinetic_avI1}-\ref{eq:variance}) requires numerical computations; however, some interesting physical insights can be achieved  by making the following analysis.

\begin{figure}
\includegraphics[width=1.\columnwidth]{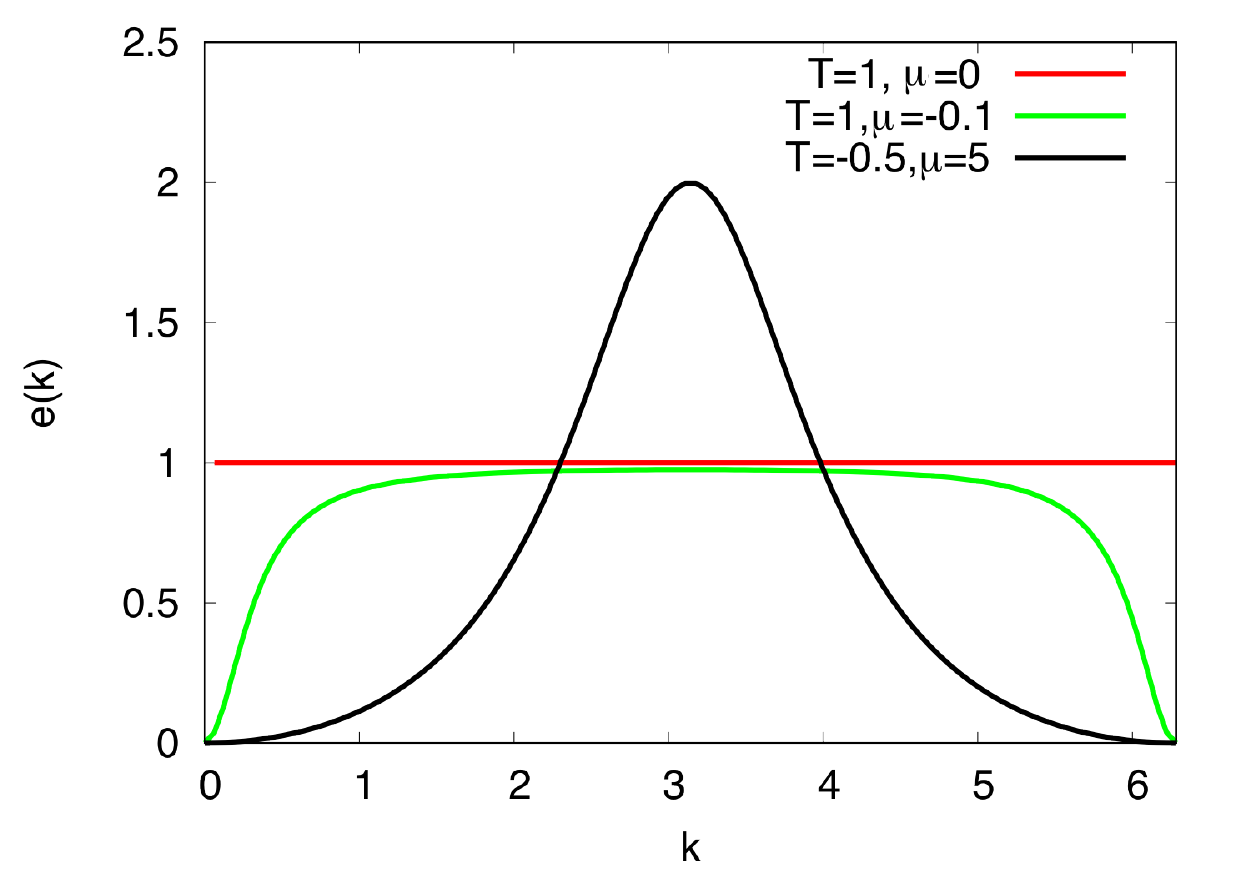}
\caption{Spectral energy density $e(k)=\omega(k) n(k)$ for different temperatures and chemical potentials. The red horizontal line is the typical equipartition of energy and corresponds to $T=1$ and $\mu=0$; the green line corresponds to $T=1$ and $\mu=0.1$ and the light blue line (sharply peaked around $k=\pi$) corresponds to negative temperatures, i.e.  $T=-0.5$ and $\mu=5$. All these curves are exact stationary solutions of the WK equation.}
\label{fig:RJ}
\end{figure}

The WK equation for phonons has two invariants:
\begin{equation}
E=\int_{0}^{2\pi} \omega(k)n(k,t) dk,\;\;\;\; N=\int_{0}^{2\pi} n(k,t) dk,
\label{eq:conserved}
\end{equation}
which are named energy and number of particles or wave action (strictly speaking, energy and number densities). Here, we point out that the conserved quantities of the  WK equation have a counterpart in the  DNLS equation; however, there is a major difference: while the number of particles is conserved in both models, the energy conserved by the WK equation corresponds to the harmonic part of the Hamiltonian, see eq. (\ref{eq:ham_fourier}), appropriately averaged. By defining the entropy density (see \cite{zakharov2012kolmogorov,spohn2006phonon}), as:
\begin{equation}
S(t)=\int_{0}^{2\pi} \ln n(k,t) dk,
\label{eq:entropy}
\end{equation}
 an $H$ theorem, $d S/dt \ge0$, holds.  When, and only when
 $dS/d t=0$,  the Rayleigh-Jeans (RJ) distribution is obtained:
\begin{equation}
n(k)^{(RJ)}=\frac{T}{\omega(k)-\mu}=\frac{1}{\beta \omega(k)-\gamma}.
\label{eq:RJ}
\end{equation}
Here  $T$ and $\mu$ are usually named temperature and  chemical potential
($\beta=1/T$ and $\gamma=\mu/T$).   This result is consistent with classical equilibrium statistical mechanics, see \cite{buonsante2016dispute}, eq. (58) therein). Because $n(k)$ is positive for all $k$, at equilibrium one of the following conditions holds: 
\begin{equation}
\begin{split}
&\beta>0 \;\;\;\; {\rm and}\;\;\; \gamma<0,\;\; {\rm or}\\
&\beta<0 \;\;\;\; {\rm and}\;\;\; \gamma<4 \beta.
\label{eq:beta}
\end{split}
\end{equation} 
As noted in \cite{buonsante2016dispute}, the last condition implies the existence of negative temperatures.
In Figure \ref{fig:RJ} we show the spectral energy density, $e(k)=\omega(k) n(k)$, as a function of $k$ for different temperatures and chemical potentials. The classical equipartition of energy, typical of systems that conserve only energy, is obtained by setting $\mu=0$. In  Figure \ref{fig:RJ}, a stationary state with $T<0$ and $\mu>0$ is also displayed. All of the states represented in Figure \ref{fig:RJ} are stationary states of the WK equation.  

%Note that in the limit of  $T\rightarrow 0^+$ and $\mu\rightarrow 0^-$, with $\mu \sim T^2$, the Rayleigh-Jeans distribution tends to a Dirac-$\delta(k)$, that implies a condensation of particles at mode  $k=0$; a similar argument leads to a condensation at $k=\pi$ in the limit of $T\rightarrow 0^-$ and $\mu\rightarrow 4^+$. 
Negative temperatures are characterized by a peaked distribution around $k=\pi$. 

Besides mean values, the wave kinetic approach offers  the possibility to investigate the fluctuations and their relaxation time scale. It can be checked by substitution that a non stationary  solution of equation (\ref{eq:variance}) is 
$\Lambda_k(t)=2 n_k^2(t)$, provided $n_k(t)$ evolves according to  (\ref{eq:discretekinetic_avI1}). The understanding of the time scale by which such solution is approached is extremely interesting.  Because of their similar mathematical structure, one may expect that equations  (\ref{eq:discretekinetic_avI1}) and (\ref{eq:variance}) evolve on the same time scale. As a matter of fact, it will be shown in the numerical computations that  $\Lambda_k(t)$ approaches  $2 n_k^2(t)$ on a much faster time scale than  the one pertaining to the evolution of $n_k$. Indeed, assuming that $\Lambda_{k}(t)$ is characterized by two time scales, the longer one being the same as the one for $n_k(t)$,  it is straightforward to show from eq. (\ref{eq:variance}) that   $\Lambda_{k}(t)$ reaches $2 n_k(t)^2$ exponentially fast and then it remains enslaved to it, as it tends to its asymptotic value.

\subsection{Equilibrium and thermodynamics}

It is not obvious a priori to what extent the variables used in  the WK equation correspond to the ones appearing  in the first law of thermodynamics. Here, we show that they satisfy the equilibrium classical relation between $T$ and $S$, i.e. $T=(\partial S/\partial E)^{-1}$.  Given 
the energy, $E$, the number of particles, $N$, and $S$ at equilibrium,
 i.e. for $n(k,t)=n(k)^{(RJ)}$, we obtain (similar integrals were calculated in \cite{rumpf2007growth} to study the erosion of a discrete breather by a thermal bath):
\begin{equation}
\begin{split}
&E(\gamma,\beta)=
\frac{2\pi}{\beta}\left(1+\frac{\gamma}{\sqrt{\gamma(\gamma-4\beta)}}\right)\,,
\\
&N(\gamma,\beta)=\frac{2\pi}{\sqrt{\gamma(\gamma-4\beta)}}\,,
\label{eq:EN_betagamma}
\end{split}
\end{equation}
\begin{equation}
\begin{split}
%S(\gamma,\beta)=\int_{0}^{2\pi} \ln n(k)^{(RJ)} dk=2\pi \ln \left[\frac{2}{2\beta -\gamma +\sqrt{\gamma(\gamma-4 \beta)}}\right]
S(\gamma,\beta)=2\pi \ln \left[\frac{2}{2\beta -\gamma +\sqrt{\gamma(\gamma-4 \beta)}}\right]\,.
\label{eq:S_gammabeta}
\end{split}
\end{equation}
To express the entropy as a function of energy and number of particles, $S(E,N)$, we invert the relations in (\ref{eq:EN_betagamma}):
\begin{equation}
\beta(E,N)=\frac{4 \pi(E-2 N)}{E (E-4 N)},\;\;\gamma(E,N)=\frac{2 \pi E}{N (E-4 N)}\,.
\label{eq:betagamma_EN}
\end{equation}
Knowing that  $\gamma=\mu/T$, the expression for the chemical potential can be derived:
\begin{equation}
\begin{split}
\mu(E,N)=\frac{E^2}{2(E-2N)N}.
\label{eq:chem_pot}
\end{split}
\end{equation}
A phase diagram with the energy as a function of number of particles for fixed temperature can be easily built by solving the first of equations (\ref{eq:betagamma_EN}) for the energy to obtain:
\begin{equation}
E=2 N +2 \pi T-2{\rm sgn}[T] \sqrt{N^2+\pi^2 T^2}.
\end{equation}
For $T\rightarrow0^+$, we have $E\rightarrow0$; for  $T\rightarrow0^-$, we have $E\rightarrow4 N$ from below; for $T \rightarrow \pm \infty$, we get $E\rightarrow2 N$.

\begin{figure}
\includegraphics[width=0.8\columnwidth]{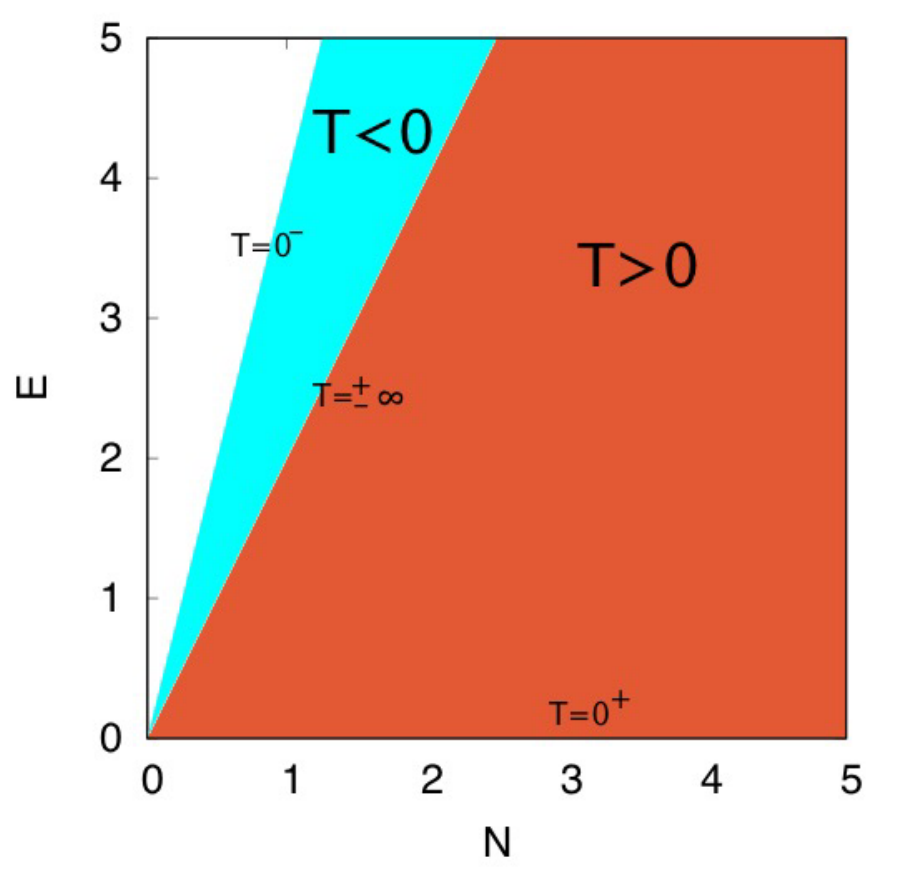}
\caption{Energy, $E$, as a function of the number of particles, $N$. The white region corresponds to non accessible energies, the light blue  to negative temperatures and the red  to positive temperatures. The lines corresponding to $0^+$, $0^-$ and $\pm \infty$ temperatures are also visible.  $T=0^-$ corresponds to $E=4 N$, $T=\pm\infty$  to $E=2 N$ and $T=0^+$  to $E=0$}
\label{fig:phase}
\end{figure}

Interestingly, since $\gamma$ is always negative, there is an upper value  for the energy for fixed number of particles, 
i.e. $0<E<4N$. Moreover,  a positive $\beta$ requires $E<2N$. Then, negative values of 
$\beta$, i.e. negative temperatures, are possibile only for $2N<E<4N$. For positive temperatures, the chemical potential is negative and becomes positive for 
negative temperatures, with the constrain that $\mu>4$. These results are shown in 
Figure \ref{fig:phase}.
Plugging 
equations (\ref{eq:betagamma_EN}) into (\ref{eq:S_gammabeta}), we obtain:
\begin{equation}
\begin{split}
S(E,N)=2\pi \ln \left[\frac{E(4 N-E)}{8\pi N}\right],
\label{eq:entropye}
\end{split}
\end{equation}
see also \cite{rumpf2009stable}.
The entropy is defined for $0 <E <4N$; it is a continuous function of its arguments and it has an absolute maximum at $E=2N$.
%, which corresponds to the chemical potential, $\mu$, equal to 0. 
For fixed $E$ and large $N$, there is a horizontal asymptote at $N=2\pi \ln [E/(2\pi)]$ which corresponds to $\gamma=0$.
\begin{figure}
\includegraphics[width=0.8\columnwidth]{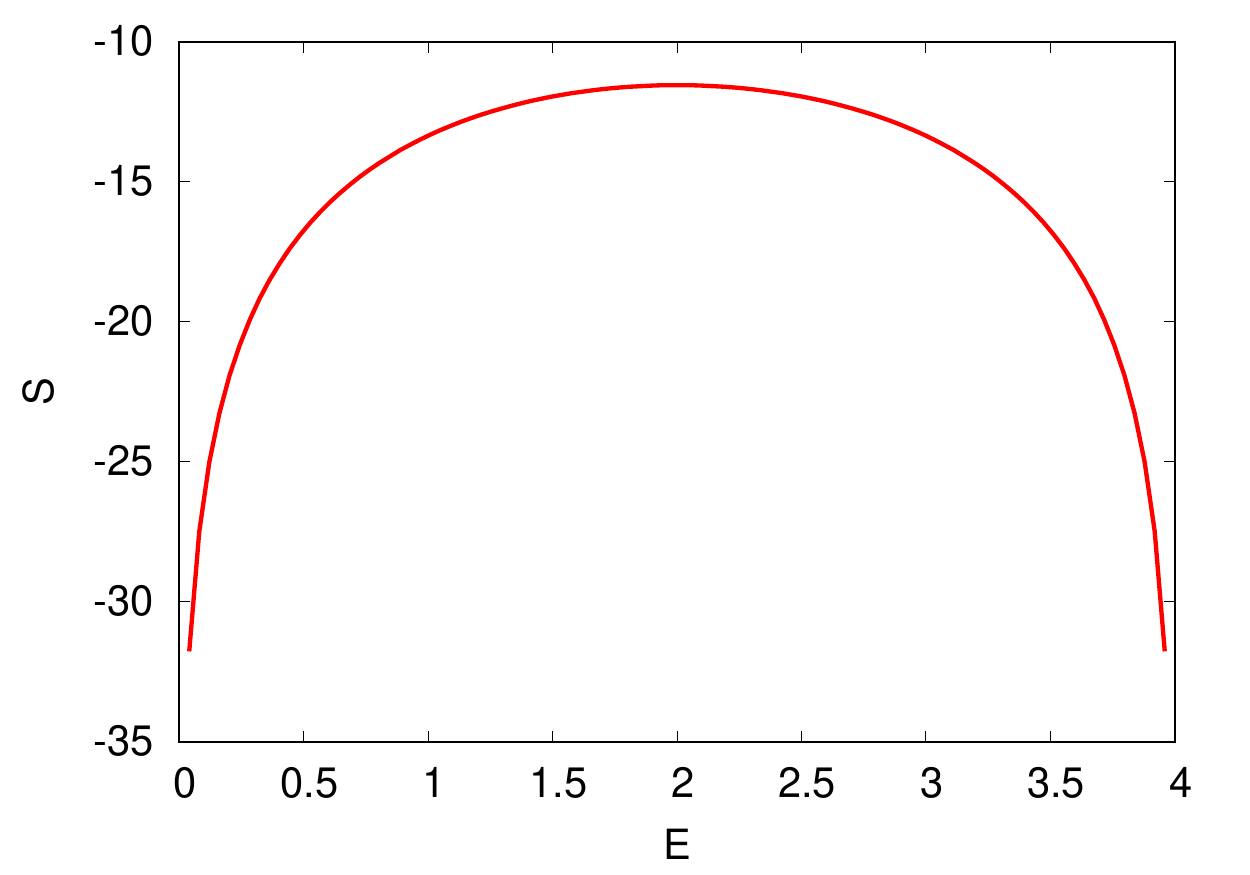}
\caption{The entropy, $S$, as a function of the energy, $E$, for $N=1$. The derivative of $S$ is the inverse of the temperature. For $E>2 N$=2 the derivative is negative, implying a negative temperature. Note that entropy is defined for $0<E<4 N$}
\label{fig:entropy}
\end{figure}
In Figure (\ref{fig:entropy}), we show the entropy as a function of the energy $E$ for $N=1$.

The above description is consistent with the formalism of classical thermodynamics; indeed, differentiating the entropy 
\begin{equation}
dS(E,N)=\left(\frac{\partial S}{\partial E}\right)_NdE+\left(\frac{\partial S}{\partial N}\right)_EdN
\end{equation}
and calculating the derivatives, it turns out that $(\partial S/ \partial E)_N=\beta$
and  $(\partial S/ \partial N)_E=-\gamma$.
This implies that the definition of entropy and other variables in the WK equation at the Rayleigh-Jeans equilibrium match the corresponding definitions 
given in classical thermodynamics.

%(the results are consistent with those contained in \cite{rumpf2004simple}, once an irrelevant shift in the energy is performed.)

%
\subsection{The Boltzmann entropy and its relation to the nonequilibrium entropy defined in equation (\ref{eq:entropy})}
 The Boltzmann entropy $S_B$ is proportional to the natural logarithm of the number of possible microstates  $\Omega$ of a system at fixed energy and number of particles:
\begin{equation}
    S_B = \ln \Omega.
    \label{entropy_bol}
\end{equation}
%Boltzmann's entropy describes the system when all the accessible microstates compatible with a macrostates are equally likely. 
%$\Omega$ is a 	function of thermodynamic variables because $S$ can be expressed as a function of temperature and chemical potential or energy and wave action.
 The strategy to compute $\Omega$ is the following: we consider  $\Omega(N,H)$
and take a two-dimensional Laplace transform to get $\Omega(\gamma,\beta)$:
\begin{equation}
\Omega(\gamma,\beta)=\int_0^{\infty} \Omega(N,H)e^{\gamma N-\beta H} dH dN
 \label{omega}
\end{equation}
$\Omega(N,H)$ can be calculated as
\begin{equation}
 \Omega(N,E)=\int_0^{\infty} \delta\left(N-\sum_{k=1}^M I_k\right)
 \delta\left(E-\sum_{k=1}^M\omega_k I_k\right)
 \prod_{k=1}^{M} d I_k,
 \label{omega_AE}
\end{equation}
where, consistently with our hypothesis related to the random phases and the smallness of the nonlinearity, we have assumed $H\simeq E$,
with $E$ the harmonic energy density. 
We plug  (\ref{omega_AE}) in (\ref{omega})  and use the property of the $\delta$, so that
\begin{equation}
\Omega(\gamma,\beta)=\int_0^{\infty} 
e^{\gamma \sum_{k=1}^M I_k -\beta \sum_{k=1}^M \omega_kI_k} 
 \prod_{k=1}^{M} d I_k,
 \label{omega1}
\end{equation}
which can be rewritten as:
\begin{equation}
\Omega(\gamma,\beta)=\int_0^{\infty} 
 \prod_{k=1}^{M}  e^{\gamma I_k -\beta  \omega_kI_k} 
d I_k .
 \label{omega1}
\end{equation}
The dependence on $I_k$ has been factorized and we can integrate over $I_k$ to get
\begin{equation}
\Omega(\gamma,\beta)=
 \prod_{k=1}^{M}  \frac{1}{-\gamma+\beta \omega_k}.
 \label{omega2}
\end{equation}
We then play the usual trick of taking the exponential of a log
\begin{equation}
\Omega(\gamma,\beta)=\exp\left[\ln
 \prod_{k=1}^{M}  \frac{1}{-\gamma+\beta \omega_k}\right]=\exp\left[\sum_{k=1}^{M}
\ln   \frac{1}{-\gamma+\beta \omega_k}\right]
 \label{omega2}
\end{equation}
Now, we take the large box limit $M=2\pi/\Delta k\rightarrow \infty$ and 
using the definition of the Boltzmann's entropy $S_B=\ln \Omega$ we get:
\begin{equation}
S_B(\gamma,\beta)=\ln \Omega(\gamma,\beta)=\frac{M}{2\pi}
\int_{0}^{2\pi}
\ln \left[ \frac{1}{-\gamma+\beta \omega_k} \right] dk
 \label{omega2}
\end{equation}
This formula, apart from the factor $M/2\pi$, is our entropy, see equation (11) where $n(k)$ has been taken at equilibrium.

%and if they are characterised by particular statistical properties of the system, such as heavy tails of the distribution of the intensities as those displayed in \cite{rasmussen2000statistical,johansson2004statistical,rumpf2008transition}

\section{Direct numerical simulations of the DNLS equation.}
The fact that the WK equation predicts the existence of negative temperatures does not necessarily imply that the DNLS equation at small nonlinearity
displays stationary solutions with $T<0$, as the WK equation is formally derived only in the limit of random phases and random amplitudes. A direct numerical simulation of the deterministic equation of motion is needed in order to establish whether the stationary solutions  of equations (\ref{eq:discretekinetic_avI1}) and (\ref{eq:variance}) are compatible with the microscopic dynamics.

\begin{figure}
\includegraphics[width=1.\columnwidth]{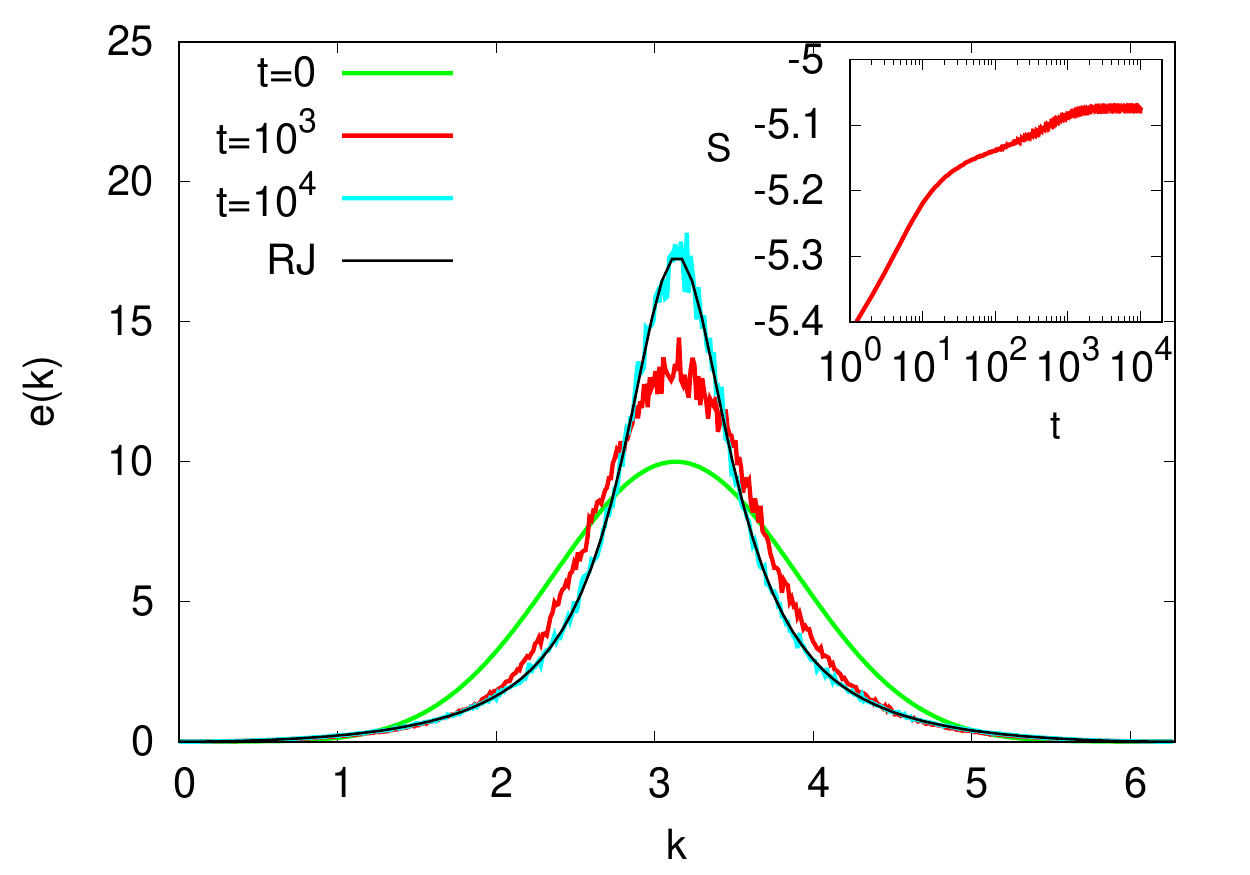}
\caption{Energy density as a function of wave number for a simulation of the DNLS equation characterized by the initial condition in (\ref{eq:init_cond}) that corresponds to  $T=-0.74$ and $\mu=4.16$. Note that, because of the conservation of energy and number of particles, temperature and chemical potential do not change in time~\cite{franzosi2011microcanonical,levy2018equilibrium}. The energy spectral density is the result of averaging over 1000 realizations characterized by different random phases. At the center of the domain, from bottom to top, the curves refer to  $t=0 $, $t=10^3$, $t=10^4$, respectively, and the dark curve is the prediction from the Rayleigh-Jeans distribution, equation (\ref{eq:RJ}). In the inset the entropy defined in eq. (\ref{eq:entropy}) as a function of time is displayed.}
\label{fig:neg_temp}
\end{figure}

\begin{figure}
\includegraphics[width=1.\columnwidth]{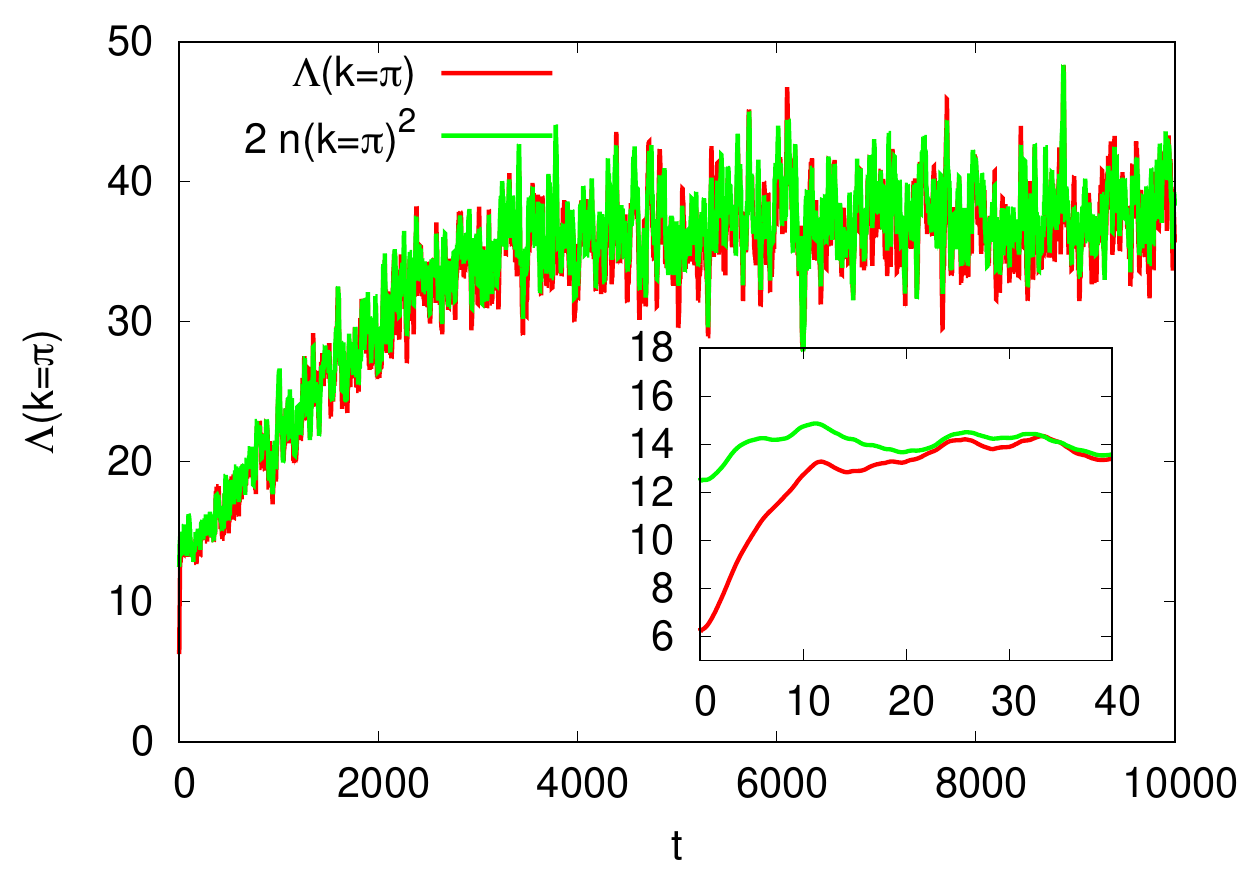}
\caption{ Evolution in time of the second moment $\Lambda(k=\pi,t)$ in light green. The solution 
of the equation (\ref{eq:variance}) $\Lambda(k=\pi,t)=2 n(k=\pi,t)^2$ is shown in red. In the inset a zoom of the the early stages of the evolution are also reported.   }
\label{fig:variance}
\end{figure}

\begin{figure}
\includegraphics[width=1.\columnwidth]{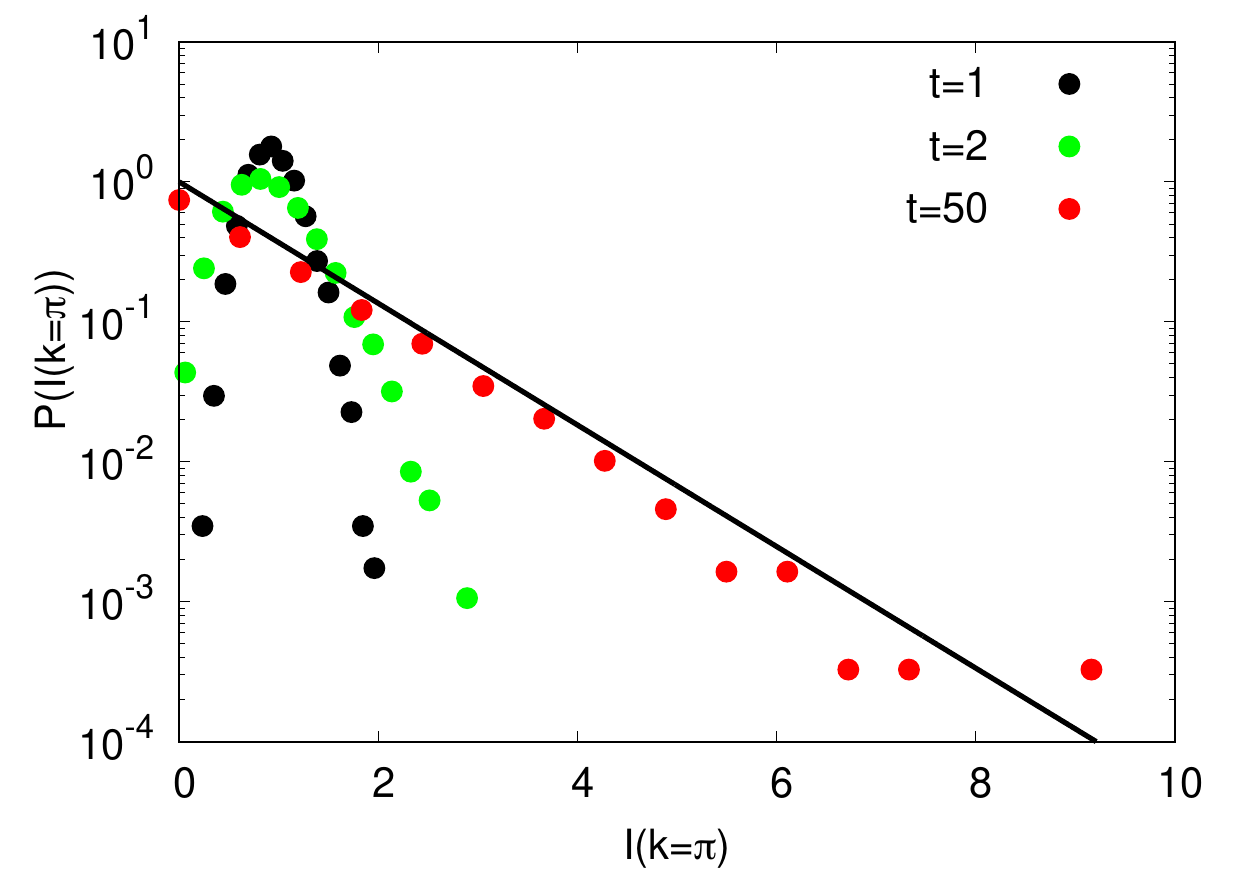}
\caption{Probability density function of $I(k=\pi)$ for different times. The dark line corresponds to the exponential distribution derived in 
Refs.\cite{chibbaro20184,chibbaro2017wave,nazarenko2011wave}}
\label{fig:pdf_intensity}
\end{figure}

The DNLS equation has been solved numerically using a standard 4th-order Runge-Kutta method; the simulations performed preserved the Hamiltonian and the number of particles  up to four significant digits. The initial conditions are provided in Fourier space; the complex amplitudes in physical space are recovered using the Discrete Fourier Transform
\begin{equation}
\psi_m=\sum_{k=1}^{M}\sqrt{n_k\Delta k}\;e^{(i 2 \pi  k m/M)} e^{i \phi_k },
\end{equation}
where $\Delta k=2\pi/M$ and $\phi_k$ are random phases distributed uniformly in the $[0,2 \pi)$ interval.
In order to observe negative temperature, we consider the following  Gaussian shaped initial wave action spectral density function:
\begin{equation}
n_k=B+A \exp\left[{\frac{-((k-k_0)\Delta k)^2}{(2 \sigma^2)}}\right]
\label{eq:init_cond}
\end{equation}
with $\sigma=0.9$, $A=2$, $B=0$, $\Delta k=2\pi/M$, $M=512$, $k_0=M/2$. With this choice 
%note that E=H_0/M and  $N=\mathrm{N}/M$
$E=\sum \omega_k n_k \Delta k=18.80$ and $N=\sum n_k \Delta k=5.63$; therefore  $2N<E<4N$ which corresponds to $T=-0.74$ and $\mu=4.16$, i.e. negative temperatures. 1000 realizations of the same spectrum (deterministic amplitudes) with different random phases have been considered and the results are obtained by averaging over the ensemble. The nonlinear parameter $\nu$ was set to 0.03. In Figure \ref{fig:neg_temp} we report the spectral energy density at time $t=0$, $t=10^3$ and  $t=10^4$; the RJ prediction with the temperature and chemical potential obtained from theory is also shown; the curves are almost indistinguishable, i.e. the long time behaviour of $e(k)$ matches the theoretical RJ predictions. The simulation was carried out up to time $t=10^5$ and no further changes in the energy density spectrum were observed  (similar results have been obtained in \cite{rumpf2008transition}). 
Moreover,  we show in the inset of the figure the monotonicity of the entropy $S$, as predicted by the $H$-theorem for the wave kinetic equation, see (\ref{eq:entropy}).
%\begin{figure}
%\includegraphics[width=0.9\columnwidth]{entropy_T<0.pdf}
%\caption{Entropy defined in eq. (\ref{eq:entropy}) as a function of time. }
%\label{fig:entropy_T<0}
%\end{figure}
Similar plots can be obtained for positive tempertures.
Concerning the fluctuations described by the second moment, we show in Figure 
 \ref{fig:variance} the
evolution in time of $\Lambda_k$ for $k=\pi$. The numerical results
show that the prediction of equation (\ref{eq:variance}) is accurate: after a very quick relaxation to 
the solution (shown in the  inset), $\Lambda_k$ follows 
the  evolution  $2n_k^2$. The probability density function of
$I_k$ is also reported in Figure \ref{fig:pdf_intensity} for different times. The prediction 
based on the wave kinetic approach is the exponential distribution \cite{chibbaro20184,chibbaro2017wave,nazarenko2011wave}; the figure shows that the distribution
tends very rapidly, on a shorter time scale than the one required for the spectrum to reach its stationary value, to the exponential curve.

\section{Discussion and conclusions} The notion of negative temperatures is  well established through some experimental results and theoretical arguments and it is well known that it is strictly connected with the existence of an upper bound for the energy. In our work we have studied negative temperatures in a lattice starting from a microscopic dynamics. The family of stationary equilibrium solutions of the WK equation associated with the lattice dynamics are characterized by two parameters which play the role of temperature and chemical potential. For most of the systems in this framework  the temperature is positive and the chemical potential is negative. However, if the dispersion relation is bounded from above, as in the case of the DNLS equation, then the distribution of particles in wave numbers can be positive also for negative temperatures and positive chemical potential. This simple observation has allowed us to carry out the calculation and express analytically $T$ and $\mu$ as  functions of the energy and number of particles. 
%The entropy, which satisfies the $H$-theorem for the WK equation, can also be found analytically at equilibrium. 
%We have verified that all the calculations are consistent with the classical thermodynamic formalism, both for positive and negative temperatures. 
Numerical simulations of the lattice dynamics agree well with the theory. Moreover, we have also studied the evolution of the fluctuations around the mean values, i.e. the wave action spectrum; in the framework of the wave kinetic approach it is also possible to derive an equation for the second moment, $\Lambda_k(t)$, of the probability density function of the  action;  it is interesting to notice that the time scale needed for $\Lambda_k(t)$ to approach its solution, $2n_k^2(t)$, is much faster than the time needed for $n_k(t)$ to reach the Rayleigh-Jeans distribution.  The probability density function of the action is shown to approach the exponential distribution on the fast time scale. 

Our analytical result is valid only in the limit of weak nonlinearity, where the unperturbed energy (the one associated with the linear part of the equation of motion) is a quasi-conserved quantity (besides the number of particles). In the presence of a single conservation law, negative temperatures are not predicted because the Rayleigh Jeans distribution corresponds to the standard equipartition of energy (no chemical potential is present) and temperature may not assume negative values.   We also emphasise that our approach, compatible with negative temperatures, is universal and  can be applied to many other dispersive wave systems characterized by resonant four-wave interactions, provided the dispersion relation is limited from above, or the Fourier domain is truncated as in the case of inviscid two dimensional turbulence, described in \cite{kraichnan1980two}. Indeed, recently, relaxation  to a RJ distribution  in a multimode optical fiber has been observed \cite{baudin2020classical}; such system is a good candidate for observing experimentally negative temperature.

\begin{acknowledgments}
M.O. and L.R. were supported by the ``Departments of Excellence 2018-2022'' Grant awarded by the Italian Ministry of
Education, University and Research (MIUR) (L.232/2016). GD acknowledges funding from ONR grant N00014-17-1-2852. M.O. was supported by Simons Collaboration on
Wave Turbulence, Grant No. 617006. B. Rumpf, S. Iubini and A. Vulpiani are acknowledged for fruitful discussions. During the writing of this manuscript, S. Nazarenko mentioned us that, in collaboration with J. Skipp, they did perform a very similar calculation on a truncated Gross-Pitaevskii equation. We acknowledge him for pointing out Ref. \cite{kraichnan1980two}.
\end{acknowledgments}

\bibliography{references}

\begin{thebibliography}{48}%
\makeatletter
\providecommand \@ifxundefined [1]{%
 \@ifx{#1\undefined}
}%
\providecommand \@ifnum [1]{%
 \ifnum #1\expandafter \@firstoftwo
 \else \expandafter \@secondoftwo
 \fi
}%
\providecommand \@ifx [1]{%
 \ifx #1\expandafter \@firstoftwo
 \else \expandafter \@secondoftwo
 \fi
}%
\providecommand \natexlab [1]{#1}%
\providecommand \enquote  [1]{``#1''}%
\providecommand \bibnamefont  [1]{#1}%
\providecommand \bibfnamefont [1]{#1}%
\providecommand \citenamefont [1]{#1}%
\providecommand \href@noop [0]{\@secondoftwo}%
\providecommand \href [0]{\begingroup \@sanitize@url \@href}%
\providecommand \@href[1]{\@@startlink{#1}\@@href}%
\providecommand \@@href[1]{\endgroup#1\@@endlink}%
\providecommand \@sanitize@url [0]{\catcode `\\12\catcode `\$12\catcode
  `\&12\catcode `\#12\catcode `\^12\catcode `\_12\catcode `\%12\relax}%
\providecommand \@@startlink[1]{}%
\providecommand \@@endlink[0]{}%
\providecommand \url  [0]{\begingroup\@sanitize@url \@url }%
\providecommand \@url [1]{\endgroup\@href {#1}{\urlprefix }}%
\providecommand \urlprefix  [0]{URL }%
\providecommand \Eprint [0]{\href }%
\providecommand \doibase [0]{https://doi.org/}%
\providecommand \selectlanguage [0]{\@gobble}%
\providecommand \bibinfo  [0]{\@secondoftwo}%
\providecommand \bibfield  [0]{\@secondoftwo}%
\providecommand \translation [1]{[#1]}%
\providecommand \BibitemOpen [0]{}%
\providecommand \bibitemStop [0]{}%
\providecommand \bibitemNoStop [0]{.\EOS\space}%
\providecommand \EOS [0]{\spacefactor3000\relax}%
\providecommand \BibitemShut  [1]{\csname bibitem#1\endcsname}%
\let\auto@bib@innerbib\@empty
%</preamble>
\bibitem [{\citenamefont {Onsager}(1949)}]{onsager1949statistical}%
  \BibitemOpen
  \bibfield  {author} {\bibinfo {author} {\bibfnamefont {L.}~\bibnamefont
  {Onsager}},\ }\bibfield  {title} {\bibinfo {title} {Statistical
  hydrodynamics},\ }\href@noop {} {\bibfield  {journal} {\bibinfo  {journal}
  {Il Nuovo Cimento (1943-1954)}\ }\textbf {\bibinfo {volume} {6}},\ \bibinfo
  {pages} {279} (\bibinfo {year} {1949})}\BibitemShut {NoStop}%
\bibitem [{\citenamefont {Purcell}\ and\ \citenamefont
  {Pound}(1951)}]{purcell1951nuclear}%
  \BibitemOpen
  \bibfield  {author} {\bibinfo {author} {\bibfnamefont {E.~M.}\ \bibnamefont
  {Purcell}}\ and\ \bibinfo {author} {\bibfnamefont {R.~V.}\ \bibnamefont
  {Pound}},\ }\bibfield  {title} {\bibinfo {title} {A nuclear spin system at
  negative temperature},\ }\href@noop {} {\bibfield  {journal} {\bibinfo
  {journal} {Physical Review}\ }\textbf {\bibinfo {volume} {81}},\ \bibinfo
  {pages} {279} (\bibinfo {year} {1951})}\BibitemShut {NoStop}%
\bibitem [{\citenamefont {Braun}\ \emph {et~al.}(2013)\citenamefont {Braun},
  \citenamefont {Ronzheimer}, \citenamefont {Schreiber}, \citenamefont
  {Hodgman}, \citenamefont {Rom}, \citenamefont {Bloch},\ and\ \citenamefont
  {Schneider}}]{braun2013negative}%
  \BibitemOpen
  \bibfield  {author} {\bibinfo {author} {\bibfnamefont {S.}~\bibnamefont
  {Braun}}, \bibinfo {author} {\bibfnamefont {J.~P.}\ \bibnamefont
  {Ronzheimer}}, \bibinfo {author} {\bibfnamefont {M.}~\bibnamefont
  {Schreiber}}, \bibinfo {author} {\bibfnamefont {S.~S.}\ \bibnamefont
  {Hodgman}}, \bibinfo {author} {\bibfnamefont {T.}~\bibnamefont {Rom}},
  \bibinfo {author} {\bibfnamefont {I.}~\bibnamefont {Bloch}},\ and\ \bibinfo
  {author} {\bibfnamefont {U.}~\bibnamefont {Schneider}},\ }\bibfield  {title}
  {\bibinfo {title} {Negative absolute temperature for motional degrees of
  freedom},\ }\href@noop {} {\bibfield  {journal} {\bibinfo  {journal}
  {Science}\ }\textbf {\bibinfo {volume} {339}},\ \bibinfo {pages} {52}
  (\bibinfo {year} {2013})}\BibitemShut {NoStop}%
\bibitem [{\citenamefont {Johnstone}\ \emph {et~al.}(2019)\citenamefont
  {Johnstone}, \citenamefont {Groszek}, \citenamefont {Starkey}, \citenamefont
  {Billington}, \citenamefont {Simula},\ and\ \citenamefont
  {Helmerson}}]{johnstone2019evolution}%
  \BibitemOpen
  \bibfield  {author} {\bibinfo {author} {\bibfnamefont {S.~P.}\ \bibnamefont
  {Johnstone}}, \bibinfo {author} {\bibfnamefont {A.~J.}\ \bibnamefont
  {Groszek}}, \bibinfo {author} {\bibfnamefont {P.~T.}\ \bibnamefont
  {Starkey}}, \bibinfo {author} {\bibfnamefont {C.~J.}\ \bibnamefont
  {Billington}}, \bibinfo {author} {\bibfnamefont {T.~P.}\ \bibnamefont
  {Simula}},\ and\ \bibinfo {author} {\bibfnamefont {K.}~\bibnamefont
  {Helmerson}},\ }\bibfield  {title} {\bibinfo {title} {Evolution of
  large-scale flow from turbulence in a two-dimensional superfluid},\
  }\href@noop {} {\bibfield  {journal} {\bibinfo  {journal} {Science}\ }\textbf
  {\bibinfo {volume} {364}},\ \bibinfo {pages} {1267} (\bibinfo {year}
  {2019})}\BibitemShut {NoStop}%
\bibitem [{\citenamefont {Gauthier}\ \emph {et~al.}(2019)\citenamefont
  {Gauthier}, \citenamefont {Reeves}, \citenamefont {Yu}, \citenamefont
  {Bradley}, \citenamefont {Baker}, \citenamefont {Bell}, \citenamefont
  {Rubinsztein-Dunlop}, \citenamefont {Davis},\ and\ \citenamefont
  {Neely}}]{gauthier2019giant}%
  \BibitemOpen
  \bibfield  {author} {\bibinfo {author} {\bibfnamefont {G.}~\bibnamefont
  {Gauthier}}, \bibinfo {author} {\bibfnamefont {M.~T.}\ \bibnamefont
  {Reeves}}, \bibinfo {author} {\bibfnamefont {X.}~\bibnamefont {Yu}}, \bibinfo
  {author} {\bibfnamefont {A.~S.}\ \bibnamefont {Bradley}}, \bibinfo {author}
  {\bibfnamefont {M.~A.}\ \bibnamefont {Baker}}, \bibinfo {author}
  {\bibfnamefont {T.~A.}\ \bibnamefont {Bell}}, \bibinfo {author}
  {\bibfnamefont {H.}~\bibnamefont {Rubinsztein-Dunlop}}, \bibinfo {author}
  {\bibfnamefont {M.~J.}\ \bibnamefont {Davis}},\ and\ \bibinfo {author}
  {\bibfnamefont {T.~W.}\ \bibnamefont {Neely}},\ }\bibfield  {title} {\bibinfo
  {title} {Giant vortex clusters in a two-dimensional quantum fluid},\
  }\href@noop {} {\bibfield  {journal} {\bibinfo  {journal} {Science}\ }\textbf
  {\bibinfo {volume} {364}},\ \bibinfo {pages} {1264} (\bibinfo {year}
  {2019})}\BibitemShut {NoStop}%
\bibitem [{\citenamefont {Ramsey}(1956)}]{ramsey1956thermodynamics}%
  \BibitemOpen
  \bibfield  {author} {\bibinfo {author} {\bibfnamefont {N.~F.}\ \bibnamefont
  {Ramsey}},\ }\bibfield  {title} {\bibinfo {title} {Thermodynamics and
  statistical mechanics at negative absolute temperatures},\ }\href@noop {}
  {\bibfield  {journal} {\bibinfo  {journal} {Physical Review}\ }\textbf
  {\bibinfo {volume} {103}},\ \bibinfo {pages} {20} (\bibinfo {year}
  {1956})}\BibitemShut {NoStop}%
\bibitem [{\citenamefont {Dunkel}\ and\ \citenamefont
  {Hilbert}(2014)}]{dunkel2014consistent}%
  \BibitemOpen
  \bibfield  {author} {\bibinfo {author} {\bibfnamefont {J.}~\bibnamefont
  {Dunkel}}\ and\ \bibinfo {author} {\bibfnamefont {S.}~\bibnamefont
  {Hilbert}},\ }\bibfield  {title} {\bibinfo {title} {Consistent
  thermostatistics forbids negative absolute temperatures},\ }\href@noop {}
  {\bibfield  {journal} {\bibinfo  {journal} {Nature Physics}\ }\textbf
  {\bibinfo {volume} {10}},\ \bibinfo {pages} {67} (\bibinfo {year}
  {2014})}\BibitemShut {NoStop}%
\bibitem [{\citenamefont {Calabrese}\ and\ \citenamefont
  {Porporato}(2019)}]{calabrese2019origin}%
  \BibitemOpen
  \bibfield  {author} {\bibinfo {author} {\bibfnamefont {S.}~\bibnamefont
  {Calabrese}}\ and\ \bibinfo {author} {\bibfnamefont {A.}~\bibnamefont
  {Porporato}},\ }\bibfield  {title} {\bibinfo {title} {Origin of negative
  temperatures in systems interacting with external fields},\ }\href@noop {}
  {\bibfield  {journal} {\bibinfo  {journal} {Physics Letters A}\ }\textbf
  {\bibinfo {volume} {383}},\ \bibinfo {pages} {2153} (\bibinfo {year}
  {2019})}\BibitemShut {NoStop}%
\bibitem [{\citenamefont {Frenkel}\ and\ \citenamefont
  {Warren}(2015)}]{frenkel2015gibbs}%
  \BibitemOpen
  \bibfield  {author} {\bibinfo {author} {\bibfnamefont {D.}~\bibnamefont
  {Frenkel}}\ and\ \bibinfo {author} {\bibfnamefont {P.~B.}\ \bibnamefont
  {Warren}},\ }\bibfield  {title} {\bibinfo {title} {Gibbs, boltzmann, and
  negative temperatures},\ }\href@noop {} {\bibfield  {journal} {\bibinfo
  {journal} {American Journal of Physics}\ }\textbf {\bibinfo {volume} {83}},\
  \bibinfo {pages} {163} (\bibinfo {year} {2015})}\BibitemShut {NoStop}%
\bibitem [{\citenamefont {Buonsante}\ \emph {et~al.}(2016)\citenamefont
  {Buonsante}, \citenamefont {Franzosi},\ and\ \citenamefont
  {Smerzi}}]{buonsante2016dispute}%
  \BibitemOpen
  \bibfield  {author} {\bibinfo {author} {\bibfnamefont {P.}~\bibnamefont
  {Buonsante}}, \bibinfo {author} {\bibfnamefont {R.}~\bibnamefont
  {Franzosi}},\ and\ \bibinfo {author} {\bibfnamefont {A.}~\bibnamefont
  {Smerzi}},\ }\bibfield  {title} {\bibinfo {title} {On the dispute between
  boltzmann and gibbs entropy},\ }\href@noop {} {\bibfield  {journal} {\bibinfo
   {journal} {Annals of Physics}\ }\textbf {\bibinfo {volume} {375}},\ \bibinfo
  {pages} {414} (\bibinfo {year} {2016})}\BibitemShut {NoStop}%
\bibitem [{\citenamefont {Puglisi}\ \emph {et~al.}(2017)\citenamefont
  {Puglisi}, \citenamefont {Sarracino},\ and\ \citenamefont
  {Vulpiani}}]{puglisi2017temperature}%
  \BibitemOpen
  \bibfield  {author} {\bibinfo {author} {\bibfnamefont {A.}~\bibnamefont
  {Puglisi}}, \bibinfo {author} {\bibfnamefont {A.}~\bibnamefont {Sarracino}},\
  and\ \bibinfo {author} {\bibfnamefont {A.}~\bibnamefont {Vulpiani}},\
  }\bibfield  {title} {\bibinfo {title} {Temperature in and out of equilibrium:
  A review of concepts, tools and attempts},\ }\href@noop {} {\bibfield
  {journal} {\bibinfo  {journal} {Physics Reports}\ }\textbf {\bibinfo {volume}
  {709}},\ \bibinfo {pages} {1} (\bibinfo {year} {2017})}\BibitemShut {NoStop}%
\bibitem [{\citenamefont {Cerino}\ \emph {et~al.}(2015)\citenamefont {Cerino},
  \citenamefont {Puglisi},\ and\ \citenamefont
  {Vulpiani}}]{cerino2015consistent}%
  \BibitemOpen
  \bibfield  {author} {\bibinfo {author} {\bibfnamefont {L.}~\bibnamefont
  {Cerino}}, \bibinfo {author} {\bibfnamefont {A.}~\bibnamefont {Puglisi}},\
  and\ \bibinfo {author} {\bibfnamefont {A.}~\bibnamefont {Vulpiani}},\
  }\bibfield  {title} {\bibinfo {title} {A consistent description of
  fluctuations requires negative temperatures},\ }\href@noop {} {\bibfield
  {journal} {\bibinfo  {journal} {Journal of Statistical Mechanics: Theory and
  Experiment}\ }\textbf {\bibinfo {volume} {2015}},\ \bibinfo {pages} {P12002}
  (\bibinfo {year} {2015})}\BibitemShut {NoStop}%
\bibitem [{\citenamefont {Abraham}\ and\ \citenamefont
  {Penrose}(2017)}]{abraham2017physics}%
  \BibitemOpen
  \bibfield  {author} {\bibinfo {author} {\bibfnamefont {E.}~\bibnamefont
  {Abraham}}\ and\ \bibinfo {author} {\bibfnamefont {O.}~\bibnamefont
  {Penrose}},\ }\bibfield  {title} {\bibinfo {title} {Physics of negative
  absolute temperatures},\ }\href@noop {} {\bibfield  {journal} {\bibinfo
  {journal} {Physical Review E}\ }\textbf {\bibinfo {volume} {95}},\ \bibinfo
  {pages} {012125} (\bibinfo {year} {2017})}\BibitemShut {NoStop}%
\bibitem [{\citenamefont {Baldovin}\ \emph {et~al.}(2021)\citenamefont
  {Baldovin}, \citenamefont {Iubini}, \citenamefont {Livi},\ and\ \citenamefont
  {Vulpiani}}]{baldovin2021statistical}%
  \BibitemOpen
  \bibfield  {author} {\bibinfo {author} {\bibfnamefont {M.}~\bibnamefont
  {Baldovin}}, \bibinfo {author} {\bibfnamefont {S.}~\bibnamefont {Iubini}},
  \bibinfo {author} {\bibfnamefont {R.}~\bibnamefont {Livi}},\ and\ \bibinfo
  {author} {\bibfnamefont {A.}~\bibnamefont {Vulpiani}},\ }\bibfield  {title}
  {\bibinfo {title} {Statistical mechanics of systems with negative
  temperature},\ }\href@noop {} {\bibfield  {journal} {\bibinfo  {journal}
  {Physics Reports}\ } (\bibinfo {year} {2021})}\BibitemShut {NoStop}%
\bibitem [{\citenamefont {Nazarenko}(2011)}]{nazarenko2011wave}%
  \BibitemOpen
  \bibfield  {author} {\bibinfo {author} {\bibfnamefont {S.}~\bibnamefont
  {Nazarenko}},\ }\href@noop {} {\emph {\bibinfo {title} {Wave turbulence}}},\
  Vol.\ \bibinfo {volume} {825}\ (\bibinfo  {publisher} {Springer Science \&
  Business Media},\ \bibinfo {year} {2011})\BibitemShut {NoStop}%
\bibitem [{\citenamefont {Zakharov}\ \emph {et~al.}(2012)\citenamefont
  {Zakharov}, \citenamefont {L'vov},\ and\ \citenamefont
  {Falkovich}}]{zakharov2012kolmogorov}%
  \BibitemOpen
  \bibfield  {author} {\bibinfo {author} {\bibfnamefont {V.~E.}\ \bibnamefont
  {Zakharov}}, \bibinfo {author} {\bibfnamefont {V.~S.}\ \bibnamefont
  {L'vov}},\ and\ \bibinfo {author} {\bibfnamefont {G.}~\bibnamefont
  {Falkovich}},\ }\href@noop {} {\emph {\bibinfo {title} {Kolmogorov spectra of
  turbulence I: Wave turbulence}}}\ (\bibinfo  {publisher} {Springer Science \&
  Business Media},\ \bibinfo {year} {2012})\BibitemShut {NoStop}%
\bibitem [{\citenamefont {Onorato}\ and\ \citenamefont
  {Dematteis}(2020)}]{onorato2020astraighforward}%
  \BibitemOpen
  \bibfield  {author} {\bibinfo {author} {\bibfnamefont {M.}~\bibnamefont
  {Onorato}}\ and\ \bibinfo {author} {\bibfnamefont {G.}~\bibnamefont
  {Dematteis}},\ }\bibfield  {title} {\bibinfo {title} {A straightforward
  derivation of the four-wave kinetic equation in action-angle variables},\
  }\href {http://iopscience.iop.org/10.1088/2399-6528/abb4b7} {\bibfield
  {journal} {\bibinfo  {journal} {Journal of Physics Communications}\ }
  (\bibinfo {year} {2020})}\BibitemShut {NoStop}%
\bibitem [{\citenamefont {Picozzi}\ \emph {et~al.}(2014)\citenamefont
  {Picozzi}, \citenamefont {Garnier}, \citenamefont {Hansson}, \citenamefont
  {Suret}, \citenamefont {Randoux}, \citenamefont {Millot},\ and\ \citenamefont
  {Christodoulides}}]{picozzi2014optical}%
  \BibitemOpen
  \bibfield  {author} {\bibinfo {author} {\bibfnamefont {A.}~\bibnamefont
  {Picozzi}}, \bibinfo {author} {\bibfnamefont {J.}~\bibnamefont {Garnier}},
  \bibinfo {author} {\bibfnamefont {T.}~\bibnamefont {Hansson}}, \bibinfo
  {author} {\bibfnamefont {P.}~\bibnamefont {Suret}}, \bibinfo {author}
  {\bibfnamefont {S.}~\bibnamefont {Randoux}}, \bibinfo {author} {\bibfnamefont
  {G.}~\bibnamefont {Millot}},\ and\ \bibinfo {author} {\bibfnamefont {D.~N.}\
  \bibnamefont {Christodoulides}},\ }\bibfield  {title} {\bibinfo {title}
  {Optical wave turbulence: Towards a unified nonequilibrium thermodynamic
  formulation of statistical nonlinear optics},\ }\href@noop {} {\bibfield
  {journal} {\bibinfo  {journal} {Physics Reports}\ }\textbf {\bibinfo {volume}
  {542}},\ \bibinfo {pages} {1} (\bibinfo {year} {2014})}\BibitemShut {NoStop}%
\bibitem [{\citenamefont {Zakharov}\ and\ \citenamefont
  {Filonenko}(1966)}]{zakharov1966energy}%
  \BibitemOpen
  \bibfield  {author} {\bibinfo {author} {\bibfnamefont {V.~E.}\ \bibnamefont
  {Zakharov}}\ and\ \bibinfo {author} {\bibfnamefont {N.}~\bibnamefont
  {Filonenko}},\ }\bibfield  {title} {\bibinfo {title} {Energy spectrum for
  stochastic oscillations of the surface of a liquid},\ }in\ \href@noop {}
  {\emph {\bibinfo {booktitle} {Doklady Akademii Nauk}}},\ Vol.\ \bibinfo
  {volume} {170}\ (\bibinfo {organization} {Russian Academy of Sciences},\
  \bibinfo {year} {1966})\ pp.\ \bibinfo {pages} {1292--1295}\BibitemShut
  {NoStop}%
\bibitem [{\citenamefont {Hasselmann}(1962)}]{hasselmann1962non}%
  \BibitemOpen
  \bibfield  {author} {\bibinfo {author} {\bibfnamefont {K.}~\bibnamefont
  {Hasselmann}},\ }\bibfield  {title} {\bibinfo {title} {On the non-linear
  energy transfer in a gravity-wave spectrum part 1. general theory},\
  }\href@noop {} {\bibfield  {journal} {\bibinfo  {journal} {Journal of Fluid
  Mechanics}\ }\textbf {\bibinfo {volume} {12}},\ \bibinfo {pages} {481}
  (\bibinfo {year} {1962})}\BibitemShut {NoStop}%
\bibitem [{\citenamefont {Nazarenko}\ and\ \citenamefont
  {Onorato}(2006)}]{nazarenko2006wave}%
  \BibitemOpen
  \bibfield  {author} {\bibinfo {author} {\bibfnamefont {S.}~\bibnamefont
  {Nazarenko}}\ and\ \bibinfo {author} {\bibfnamefont {M.}~\bibnamefont
  {Onorato}},\ }\bibfield  {title} {\bibinfo {title} {Wave turbulence and
  vortices in bose--einstein condensation},\ }\href@noop {} {\bibfield
  {journal} {\bibinfo  {journal} {Physica D: Nonlinear Phenomena}\ }\textbf
  {\bibinfo {volume} {219}},\ \bibinfo {pages} {1} (\bibinfo {year}
  {2006})}\BibitemShut {NoStop}%
\bibitem [{\citenamefont {Proment}\ \emph {et~al.}(2009)\citenamefont
  {Proment}, \citenamefont {Nazarenko},\ and\ \citenamefont
  {Onorato}}]{proment2009quantum}%
  \BibitemOpen
  \bibfield  {author} {\bibinfo {author} {\bibfnamefont {D.}~\bibnamefont
  {Proment}}, \bibinfo {author} {\bibfnamefont {S.}~\bibnamefont {Nazarenko}},\
  and\ \bibinfo {author} {\bibfnamefont {M.}~\bibnamefont {Onorato}},\
  }\bibfield  {title} {\bibinfo {title} {Quantum turbulence cascades in the
  gross-pitaevskii model},\ }\href@noop {} {\bibfield  {journal} {\bibinfo
  {journal} {Physical Review A}\ }\textbf {\bibinfo {volume} {80}},\ \bibinfo
  {pages} {051603} (\bibinfo {year} {2009})}\BibitemShut {NoStop}%
\bibitem [{\citenamefont {Galtier}\ and\ \citenamefont
  {Nazarenko}(2017)}]{galtier2017turbulence}%
  \BibitemOpen
  \bibfield  {author} {\bibinfo {author} {\bibfnamefont {S.}~\bibnamefont
  {Galtier}}\ and\ \bibinfo {author} {\bibfnamefont {S.~V.}\ \bibnamefont
  {Nazarenko}},\ }\bibfield  {title} {\bibinfo {title} {Turbulence of weak
  gravitational waves in the early universe},\ }\href@noop {} {\bibfield
  {journal} {\bibinfo  {journal} {Physical Review Letters}\ }\textbf {\bibinfo
  {volume} {119}},\ \bibinfo {pages} {221101} (\bibinfo {year}
  {2017})}\BibitemShut {NoStop}%
\bibitem [{\citenamefont {Lukkarinen}\ and\ \citenamefont
  {Spohn}(2008)}]{lukkarinen2008anomalous}%
  \BibitemOpen
  \bibfield  {author} {\bibinfo {author} {\bibfnamefont {J.}~\bibnamefont
  {Lukkarinen}}\ and\ \bibinfo {author} {\bibfnamefont {H.}~\bibnamefont
  {Spohn}},\ }\bibfield  {title} {\bibinfo {title} {Anomalous energy transport
  in the fpu-$\beta$ chain},\ }\href@noop {} {\bibfield  {journal} {\bibinfo
  {journal} {Communications on Pure and Applied Mathematics: A Journal Issued
  by the Courant Institute of Mathematical Sciences}\ }\textbf {\bibinfo
  {volume} {61}},\ \bibinfo {pages} {1753} (\bibinfo {year}
  {2008})}\BibitemShut {NoStop}%
\bibitem [{\citenamefont {Onorato}\ \emph {et~al.}(2015)\citenamefont
  {Onorato}, \citenamefont {Vozella}, \citenamefont {Proment},\ and\
  \citenamefont {Lvov}}]{onorato2015route}%
  \BibitemOpen
  \bibfield  {author} {\bibinfo {author} {\bibfnamefont {M.}~\bibnamefont
  {Onorato}}, \bibinfo {author} {\bibfnamefont {L.}~\bibnamefont {Vozella}},
  \bibinfo {author} {\bibfnamefont {D.}~\bibnamefont {Proment}},\ and\ \bibinfo
  {author} {\bibfnamefont {Y.~V.}\ \bibnamefont {Lvov}},\ }\bibfield  {title}
  {\bibinfo {title} {Route to thermalization in the $\alpha$-fermi--pasta--ulam
  system},\ }\href@noop {} {\bibfield  {journal} {\bibinfo  {journal}
  {Proceedings of the National Academy of Sciences}\ }\textbf {\bibinfo
  {volume} {112}},\ \bibinfo {pages} {4208} (\bibinfo {year}
  {2015})}\BibitemShut {NoStop}%
\bibitem [{\citenamefont {Kevrekidis}(2009)}]{kevrekidis2009discrete}%
  \BibitemOpen
  \bibfield  {author} {\bibinfo {author} {\bibfnamefont {P.~G.}\ \bibnamefont
  {Kevrekidis}},\ }\href@noop {} {\emph {\bibinfo {title} {The discrete
  nonlinear Schr{\"o}dinger equation: mathematical analysis, numerical
  computations and physical perspectives}}},\ Vol.\ \bibinfo {volume} {232}\
  (\bibinfo  {publisher} {Springer Science \& Business Media},\ \bibinfo {year}
  {2009})\BibitemShut {NoStop}%
\bibitem [{\citenamefont {Fermi}\ \emph {et~al.}(1955)\citenamefont {Fermi},
  \citenamefont {Pasta},\ and\ \citenamefont {Ulam}}]{fermi1955alamos}%
  \BibitemOpen
  \bibfield  {author} {\bibinfo {author} {\bibfnamefont {E.}~\bibnamefont
  {Fermi}}, \bibinfo {author} {\bibfnamefont {J.}~\bibnamefont {Pasta}},\ and\
  \bibinfo {author} {\bibfnamefont {S.}~\bibnamefont {Ulam}},\ }\bibfield
  {title} {\bibinfo {title} {{L}os {A}lamos {R}eport {LA}-1940},\ }\href@noop
  {} {\bibfield  {journal} {\bibinfo  {journal} {E. Fermi, Collected Papers}\
  }\textbf {\bibinfo {volume} {2}},\ \bibinfo {pages} {977} (\bibinfo {year}
  {1955})}\BibitemShut {NoStop}%
\bibitem [{\citenamefont {Rasmussen}\ \emph {et~al.}(2000)\citenamefont
  {Rasmussen}, \citenamefont {Cretegny}, \citenamefont {Kevrekidis},\ and\
  \citenamefont {Gr{\o}nbech-Jensen}}]{rasmussen2000statistical}%
  \BibitemOpen
  \bibfield  {author} {\bibinfo {author} {\bibfnamefont {K.}~\bibnamefont
  {Rasmussen}}, \bibinfo {author} {\bibfnamefont {T.}~\bibnamefont {Cretegny}},
  \bibinfo {author} {\bibfnamefont {P.~G.}\ \bibnamefont {Kevrekidis}},\ and\
  \bibinfo {author} {\bibfnamefont {N.}~\bibnamefont {Gr{\o}nbech-Jensen}},\
  }\bibfield  {title} {\bibinfo {title} {Statistical mechanics of a discrete
  nonlinear system},\ }\href@noop {} {\bibfield  {journal} {\bibinfo  {journal}
  {Physical review letters}\ }\textbf {\bibinfo {volume} {84}},\ \bibinfo
  {pages} {3740} (\bibinfo {year} {2000})}\BibitemShut {NoStop}%
\bibitem [{\citenamefont {Rumpf}(2008)}]{rumpf2008transition}%
  \BibitemOpen
  \bibfield  {author} {\bibinfo {author} {\bibfnamefont {B.}~\bibnamefont
  {Rumpf}},\ }\bibfield  {title} {\bibinfo {title} {Transition behavior of the
  discrete nonlinear schr{\"o}dinger equation},\ }\href@noop {} {\bibfield
  {journal} {\bibinfo  {journal} {Physical Review E}\ }\textbf {\bibinfo
  {volume} {77}},\ \bibinfo {pages} {036606} (\bibinfo {year}
  {2008})}\BibitemShut {NoStop}%
\bibitem [{\citenamefont {Rumpf}(2009)}]{rumpf2009stable}%
  \BibitemOpen
  \bibfield  {author} {\bibinfo {author} {\bibfnamefont {B.}~\bibnamefont
  {Rumpf}},\ }\bibfield  {title} {\bibinfo {title} {Stable and metastable
  states and the formation and destruction of breathers in the discrete
  nonlinear schr{\"o}dinger equation},\ }\href@noop {} {\bibfield  {journal}
  {\bibinfo  {journal} {Physica D: Nonlinear Phenomena}\ }\textbf {\bibinfo
  {volume} {238}},\ \bibinfo {pages} {2067} (\bibinfo {year}
  {2009})}\BibitemShut {NoStop}%
\bibitem [{\citenamefont {Iubini}\ \emph {et~al.}(2012)\citenamefont {Iubini},
  \citenamefont {Lepri},\ and\ \citenamefont
  {Politi}}]{iubini2012nonequilibrium}%
  \BibitemOpen
  \bibfield  {author} {\bibinfo {author} {\bibfnamefont {S.}~\bibnamefont
  {Iubini}}, \bibinfo {author} {\bibfnamefont {S.}~\bibnamefont {Lepri}},\ and\
  \bibinfo {author} {\bibfnamefont {A.}~\bibnamefont {Politi}},\ }\bibfield
  {title} {\bibinfo {title} {Nonequilibrium discrete nonlinear schr{\"o}dinger
  equation},\ }\href@noop {} {\bibfield  {journal} {\bibinfo  {journal}
  {Physical Review E}\ }\textbf {\bibinfo {volume} {86}},\ \bibinfo {pages}
  {011108} (\bibinfo {year} {2012})}\BibitemShut {NoStop}%
\bibitem [{\citenamefont {Iubini}\ \emph {et~al.}(2013)\citenamefont {Iubini},
  \citenamefont {Franzosi}, \citenamefont {Livi}, \citenamefont {Oppo},\ and\
  \citenamefont {Politi}}]{iubini2013discrete}%
  \BibitemOpen
  \bibfield  {author} {\bibinfo {author} {\bibfnamefont {S.}~\bibnamefont
  {Iubini}}, \bibinfo {author} {\bibfnamefont {R.}~\bibnamefont {Franzosi}},
  \bibinfo {author} {\bibfnamefont {R.}~\bibnamefont {Livi}}, \bibinfo {author}
  {\bibfnamefont {G.-L.}\ \bibnamefont {Oppo}},\ and\ \bibinfo {author}
  {\bibfnamefont {A.}~\bibnamefont {Politi}},\ }\bibfield  {title} {\bibinfo
  {title} {Discrete breathers and negative-temperature states},\ }\href@noop {}
  {\bibfield  {journal} {\bibinfo  {journal} {New Journal of Physics}\ }\textbf
  {\bibinfo {volume} {15}},\ \bibinfo {pages} {023032} (\bibinfo {year}
  {2013})}\BibitemShut {NoStop}%
\bibitem [{\citenamefont {Rumpf}(2007)}]{rumpf2007growth}%
  \BibitemOpen
  \bibfield  {author} {\bibinfo {author} {\bibfnamefont {B.}~\bibnamefont
  {Rumpf}},\ }\bibfield  {title} {\bibinfo {title} {Growth and erosion of a
  discrete breather interacting with rayleigh-jeans distributed phonons},\
  }\href@noop {} {\bibfield  {journal} {\bibinfo  {journal} {EPL (Europhysics
  Letters)}\ }\textbf {\bibinfo {volume} {78}},\ \bibinfo {pages} {26001}
  (\bibinfo {year} {2007})}\BibitemShut {NoStop}%
\bibitem [{\citenamefont {Rumpf}(2004)}]{rumpf2004simple}%
  \BibitemOpen
  \bibfield  {author} {\bibinfo {author} {\bibfnamefont {B.}~\bibnamefont
  {Rumpf}},\ }\bibfield  {title} {\bibinfo {title} {Simple statistical
  explanation for the localization of energy in nonlinear lattices with two
  conserved quantities},\ }\href@noop {} {\bibfield  {journal} {\bibinfo
  {journal} {Physical Review E}\ }\textbf {\bibinfo {volume} {69}},\ \bibinfo
  {pages} {016618} (\bibinfo {year} {2004})}\BibitemShut {NoStop}%
\bibitem [{\citenamefont {Flach}\ and\ \citenamefont
  {Willis}(1998)}]{flach1998discrete}%
  \BibitemOpen
  \bibfield  {author} {\bibinfo {author} {\bibfnamefont {S.}~\bibnamefont
  {Flach}}\ and\ \bibinfo {author} {\bibfnamefont {C.~R.}\ \bibnamefont
  {Willis}},\ }\bibfield  {title} {\bibinfo {title} {Discrete breathers},\
  }\href@noop {} {\bibfield  {journal} {\bibinfo  {journal} {Physics Reports}\
  }\textbf {\bibinfo {volume} {295}},\ \bibinfo {pages} {181} (\bibinfo {year}
  {1998})}\BibitemShut {NoStop}%
\bibitem [{\citenamefont {Gradenigo}\ \emph {et~al.}(2019)\citenamefont
  {Gradenigo}, \citenamefont {Iubini}, \citenamefont {Livi},\ and\
  \citenamefont {Majumdar}}]{gradenigo2019localization}%
  \BibitemOpen
  \bibfield  {author} {\bibinfo {author} {\bibfnamefont {G.}~\bibnamefont
  {Gradenigo}}, \bibinfo {author} {\bibfnamefont {S.}~\bibnamefont {Iubini}},
  \bibinfo {author} {\bibfnamefont {R.}~\bibnamefont {Livi}},\ and\ \bibinfo
  {author} {\bibfnamefont {S.~N.}\ \bibnamefont {Majumdar}},\ }\href@noop {}
  {\bibinfo {title} {Localization in the discrete non-linear schr\"odinger
  equation: mechanism of a first-order transition in the microcanonical
  ensemble}} (\bibinfo {year} {2019}),\ \Eprint
  {https://arxiv.org/abs/1910.07461} {arXiv:1910.07461 [cond-mat.stat-mech]}
  \BibitemShut {NoStop}%
\bibitem [{\citenamefont {Parto}\ \emph {et~al.}(2019)\citenamefont {Parto},
  \citenamefont {Wu}, \citenamefont {Jung}, \citenamefont {Makris},\ and\
  \citenamefont {Christodoulides}}]{parto2019thermodynamic}%
  \BibitemOpen
  \bibfield  {author} {\bibinfo {author} {\bibfnamefont {M.}~\bibnamefont
  {Parto}}, \bibinfo {author} {\bibfnamefont {F.~O.}\ \bibnamefont {Wu}},
  \bibinfo {author} {\bibfnamefont {P.~S.}\ \bibnamefont {Jung}}, \bibinfo
  {author} {\bibfnamefont {K.}~\bibnamefont {Makris}},\ and\ \bibinfo {author}
  {\bibfnamefont {D.~N.}\ \bibnamefont {Christodoulides}},\ }\bibfield  {title}
  {\bibinfo {title} {Thermodynamic conditions governing the optical temperature
  and chemical potential in nonlinear highly multimoded photonic systems},\
  }\href@noop {} {\bibfield  {journal} {\bibinfo  {journal} {Optics letters}\
  }\textbf {\bibinfo {volume} {44}},\ \bibinfo {pages} {3936} (\bibinfo {year}
  {2019})}\BibitemShut {NoStop}%
\bibitem [{\citenamefont {Makris}\ \emph {et~al.}(2020)\citenamefont {Makris},
  \citenamefont {Wu}, \citenamefont {Jung},\ and\ \citenamefont
  {Christodoulides}}]{makris2020statistical}%
  \BibitemOpen
  \bibfield  {author} {\bibinfo {author} {\bibfnamefont {K.~G.}\ \bibnamefont
  {Makris}}, \bibinfo {author} {\bibfnamefont {F.~O.}\ \bibnamefont {Wu}},
  \bibinfo {author} {\bibfnamefont {P.~S.}\ \bibnamefont {Jung}},\ and\
  \bibinfo {author} {\bibfnamefont {D.~N.}\ \bibnamefont {Christodoulides}},\
  }\bibfield  {title} {\bibinfo {title} {Statistical mechanics of weakly
  nonlinear optical multimode gases},\ }\href@noop {} {\bibfield  {journal}
  {\bibinfo  {journal} {Optics Letters}\ }\textbf {\bibinfo {volume} {45}},\
  \bibinfo {pages} {1651} (\bibinfo {year} {2020})}\BibitemShut {NoStop}%
\bibitem [{\citenamefont {Wu}\ \emph {et~al.}(2019)\citenamefont {Wu},
  \citenamefont {Hassan},\ and\ \citenamefont
  {Christodoulides}}]{wu2019thermodynamic}%
  \BibitemOpen
  \bibfield  {author} {\bibinfo {author} {\bibfnamefont {F.~O.}\ \bibnamefont
  {Wu}}, \bibinfo {author} {\bibfnamefont {A.~U.}\ \bibnamefont {Hassan}},\
  and\ \bibinfo {author} {\bibfnamefont {D.~N.}\ \bibnamefont
  {Christodoulides}},\ }\bibfield  {title} {\bibinfo {title} {Thermodynamic
  theory of highly multimoded nonlinear optical systems},\ }\href@noop {}
  {\bibfield  {journal} {\bibinfo  {journal} {Nature Photonics}\ }\textbf
  {\bibinfo {volume} {13}},\ \bibinfo {pages} {776} (\bibinfo {year}
  {2019})}\BibitemShut {NoStop}%
\bibitem [{\citenamefont {Wu}\ \emph {et~al.}(2020)\citenamefont {Wu},
  \citenamefont {Jung}, \citenamefont {Parto}, \citenamefont {Khajavikhan},\
  and\ \citenamefont {Christodoulides}}]{wu2020entropic}%
  \BibitemOpen
  \bibfield  {author} {\bibinfo {author} {\bibfnamefont {F.~O.}\ \bibnamefont
  {Wu}}, \bibinfo {author} {\bibfnamefont {P.~S.}\ \bibnamefont {Jung}},
  \bibinfo {author} {\bibfnamefont {M.}~\bibnamefont {Parto}}, \bibinfo
  {author} {\bibfnamefont {M.}~\bibnamefont {Khajavikhan}},\ and\ \bibinfo
  {author} {\bibfnamefont {D.~N.}\ \bibnamefont {Christodoulides}},\ }\bibfield
   {title} {\bibinfo {title} {Entropic thermodynamics of nonlinear photonic
  chain networks},\ }\href@noop {} {\bibfield  {journal} {\bibinfo  {journal}
  {Communications Physics}\ }\textbf {\bibinfo {volume} {3}},\ \bibinfo {pages}
  {1} (\bibinfo {year} {2020})}\BibitemShut {NoStop}%
\bibitem [{\citenamefont {Chibbaro}\ \emph {et~al.}(2018)\citenamefont
  {Chibbaro}, \citenamefont {Dematteis},\ and\ \citenamefont
  {Rondoni}}]{chibbaro20184}%
  \BibitemOpen
  \bibfield  {author} {\bibinfo {author} {\bibfnamefont {S.}~\bibnamefont
  {Chibbaro}}, \bibinfo {author} {\bibfnamefont {G.}~\bibnamefont
  {Dematteis}},\ and\ \bibinfo {author} {\bibfnamefont {L.}~\bibnamefont
  {Rondoni}},\ }\bibfield  {title} {\bibinfo {title} {4-wave dynamics in
  kinetic wave turbulence},\ }\href@noop {} {\bibfield  {journal} {\bibinfo
  {journal} {Physica D: Nonlinear Phenomena}\ }\textbf {\bibinfo {volume}
  {362}},\ \bibinfo {pages} {24} (\bibinfo {year} {2018})}\BibitemShut
  {NoStop}%
\bibitem [{\citenamefont {Tanaka}\ and\ \citenamefont
  {Yokoyama}(2013)}]{tanaka2013numerical}%
  \BibitemOpen
  \bibfield  {author} {\bibinfo {author} {\bibfnamefont {M.}~\bibnamefont
  {Tanaka}}\ and\ \bibinfo {author} {\bibfnamefont {N.}~\bibnamefont
  {Yokoyama}},\ }\bibfield  {title} {\bibinfo {title} {Numerical verification
  of the random-phase-and-amplitude formalism of weak turbulence},\ }\href@noop
  {} {\bibfield  {journal} {\bibinfo  {journal} {Physical Review E}\ }\textbf
  {\bibinfo {volume} {87}},\ \bibinfo {pages} {062922} (\bibinfo {year}
  {2013})}\BibitemShut {NoStop}%
\bibitem [{\citenamefont {Spohn}(2006)}]{spohn2006phonon}%
  \BibitemOpen
  \bibfield  {author} {\bibinfo {author} {\bibfnamefont {H.}~\bibnamefont
  {Spohn}},\ }\bibfield  {title} {\bibinfo {title} {The phonon boltzmann
  equation, properties and link to weakly anharmonic lattice dynamics},\
  }\href@noop {} {\bibfield  {journal} {\bibinfo  {journal} {Journal of
  statistical physics}\ }\textbf {\bibinfo {volume} {124}},\ \bibinfo {pages}
  {1041} (\bibinfo {year} {2006})}\BibitemShut {NoStop}%
\bibitem [{\citenamefont {Franzosi}(2011)}]{franzosi2011microcanonical}%
  \BibitemOpen
  \bibfield  {author} {\bibinfo {author} {\bibfnamefont {R.}~\bibnamefont
  {Franzosi}},\ }\bibfield  {title} {\bibinfo {title} {Microcanonical entropy
  and dynamical measure of temperature for systems with two first integrals},\
  }\href@noop {} {\bibfield  {journal} {\bibinfo  {journal} {Journal of
  Statistical Physics}\ }\textbf {\bibinfo {volume} {143}},\ \bibinfo {pages}
  {824} (\bibinfo {year} {2011})}\BibitemShut {NoStop}%
\bibitem [{\citenamefont {Levy}\ and\ \citenamefont
  {Silberberg}(2018)}]{levy2018equilibrium}%
  \BibitemOpen
  \bibfield  {author} {\bibinfo {author} {\bibfnamefont {U.}~\bibnamefont
  {Levy}}\ and\ \bibinfo {author} {\bibfnamefont {Y.}~\bibnamefont
  {Silberberg}},\ }\bibfield  {title} {\bibinfo {title} {Equilibrium
  temperatures of discrete nonlinear systems},\ }\href@noop {} {\bibfield
  {journal} {\bibinfo  {journal} {Physical Review B}\ }\textbf {\bibinfo
  {volume} {98}},\ \bibinfo {pages} {060303} (\bibinfo {year}
  {2018})}\BibitemShut {NoStop}%
\bibitem [{\citenamefont {Chibbaro}\ \emph {et~al.}(2017)\citenamefont
  {Chibbaro}, \citenamefont {Dematteis}, \citenamefont {Josserand},\ and\
  \citenamefont {Rondoni}}]{chibbaro2017wave}%
  \BibitemOpen
  \bibfield  {author} {\bibinfo {author} {\bibfnamefont {S.}~\bibnamefont
  {Chibbaro}}, \bibinfo {author} {\bibfnamefont {G.}~\bibnamefont {Dematteis}},
  \bibinfo {author} {\bibfnamefont {C.}~\bibnamefont {Josserand}},\ and\
  \bibinfo {author} {\bibfnamefont {L.}~\bibnamefont {Rondoni}},\ }\bibfield
  {title} {\bibinfo {title} {Wave-turbulence theory of four-wave nonlinear
  interactions},\ }\href@noop {} {\bibfield  {journal} {\bibinfo  {journal}
  {Physical Review E}\ }\textbf {\bibinfo {volume} {96}},\ \bibinfo {pages}
  {021101} (\bibinfo {year} {2017})}\BibitemShut {NoStop}%
\bibitem [{\citenamefont {Kraichnan}\ and\ \citenamefont
  {Montgomery}(1980)}]{kraichnan1980two}%
  \BibitemOpen
  \bibfield  {author} {\bibinfo {author} {\bibfnamefont {R.~H.}\ \bibnamefont
  {Kraichnan}}\ and\ \bibinfo {author} {\bibfnamefont {D.}~\bibnamefont
  {Montgomery}},\ }\bibfield  {title} {\bibinfo {title} {Two-dimensional
  turbulence},\ }\href@noop {} {\bibfield  {journal} {\bibinfo  {journal}
  {Reports on Progress in Physics}\ }\textbf {\bibinfo {volume} {43}},\
  \bibinfo {pages} {547} (\bibinfo {year} {1980})}\BibitemShut {NoStop}%
\bibitem [{\citenamefont {Baudin}\ \emph {et~al.}(2020)\citenamefont {Baudin},
  \citenamefont {Fusaro}, \citenamefont {Krupa}, \citenamefont {Garnier},
  \citenamefont {Rica}, \citenamefont {Millot},\ and\ \citenamefont
  {Picozzi}}]{baudin2020classical}%
  \BibitemOpen
  \bibfield  {author} {\bibinfo {author} {\bibfnamefont {K.}~\bibnamefont
  {Baudin}}, \bibinfo {author} {\bibfnamefont {A.}~\bibnamefont {Fusaro}},
  \bibinfo {author} {\bibfnamefont {K.}~\bibnamefont {Krupa}}, \bibinfo
  {author} {\bibfnamefont {J.}~\bibnamefont {Garnier}}, \bibinfo {author}
  {\bibfnamefont {S.}~\bibnamefont {Rica}}, \bibinfo {author} {\bibfnamefont
  {G.}~\bibnamefont {Millot}},\ and\ \bibinfo {author} {\bibfnamefont
  {A.}~\bibnamefont {Picozzi}},\ }\bibfield  {title} {\bibinfo {title}
  {Classical rayleigh-jeans condensation of light waves: Observation and
  thermodynamic characterization},\ }\href@noop {} {\bibfield  {journal}
  {\bibinfo  {journal} {Physical Review Letters}\ }\textbf {\bibinfo {volume}
  {125}},\ \bibinfo {pages} {244101} (\bibinfo {year} {2020})}\BibitemShut
  {NoStop}%
\end{thebibliography}%

\end{document}